
\font\titlefont = cmr10 scaled \magstep2
\magnification=\magstep1
\vsize=22truecm
\voffset=1.75truecm
\hsize=15truecm
\hoffset=0.95truecm
\baselineskip=20pt

\settabs 18 \columns

\def\b{\bigskip}
\def\bb{\bigskip\bigskip}

\def\ce{\centerline}

\def\no{\noindent}

\rightline{AMES-HET-94-03}
\rightline{May 2, 1994}
\bb

\ce{\titlefont {The Chiral Lagrangian Approach To}}
\ce{\titlefont{The Non-Universal Effects
  In The $Z b {\overline b}$ Vertex}}

\bb

\b

\ce{\bf {X. Zhang} }
\b
\ce{\it{ Department of Physics and Astronomy} }
\ce{\it {Iowa State University}}
\ce{\it{ Ames, Iowa
 50011}}

\bb
\b

\b
\bb
\ce{\bf ABSTRACT}
\b
\no We calculate the top quark loop corrections to
the $Z b {\overline b}$ vertex in
 the chiral lagrangian of the dynamical electroweak
symmetry breaking theory. We use the
effective lagrangian based on coset
 space $SU(2) \times U(1) / U(1)$ with the electroweak
gauge interaction switched off, so these corrections are
directly related to the physics associated with the heavy top quark and
the mass generation in the dynamical models.

\bb
\filbreak

The universality of the weak interaction is argued[1] to be violated
in the dynamical electroweak $SU_L(2) \times U_Y(1)$ symmetry
breaking theory. These non-universal gauge
interactions, in terms of the Goldberger-Treiman relation[2],
 are directly related to the
 non-universal couplings of the matter fields to the Goldstone bosons
 associated with the longitudinal weak gauge bosons.
Thus one can study these non-universal effects in the
 chiral lagrangian with gauge couplings turned off.
In this paper we will make a detail analysis of the non-universal effects
on the $Z b {\overline b }$ vertex since LEP data show that the
$Z b {\overline b}$ partial width lies slightly above the standard
model prediction[3]. Specifically we will calculate the
top quark loop corrections
to the interaction
 of the bottom quark with the neutral Goldstone (longitudinal
Z) boson.

First of all, let us examine the interactions among Goldstone bosons
and fermions in the standard model.
 By turning off the electroweak gauge interactions, there is a
global $SU_L(2) \times U_Y(1)$ in the standard model.
For our purpose, let us restrict ourself to bottom and top quarks
in the fermion sector. The lagrangian here is:
$${
{\cal L} =
{\cal L}_1  +  {\cal L}_2  + {\cal L}_3
{}~~~, } \eqno(1)$$
\no where

$${
{\cal L}_1 = {\overline F_L} i \gamma_\mu
            \partial^\mu F_L
 +{\overline t^\prime}_R i \gamma^\mu
 \partial_\mu t_R^\prime + {\overline b^\prime}_R i \gamma^\mu
                 \partial_\mu b_R^\prime ~~; }$$

$${
{\cal L}_2 = \partial^\mu \Phi^\dagger \partial_\mu \Phi - V( \Phi )
{}~~~~; }$$

$${
{\cal L}_3 = f^t ~{\overline F_L} {\tilde\Phi}
               t_R^\prime  + f^b ~ {\overline F_L}
           \Phi b_R^\prime + h.c
{}~~~~; }$$
\no and

$${  F_L = {\pmatrix{ t^\prime \cr b^\prime \cr } }_L , ~~
{\tilde\Phi} = i \tau_2 \Phi^* ~, ~~{\rm and} ~~
          V( \Phi ) = {\lambda \over 4} {( \Phi^\dagger \Phi - {v^2 \over 2}
                     ) }^2  ~~~. }$$

\no As $\Phi$ develops a vacuum expectation value due to the
 asymmetric potential $V( \Phi )$,
the electroweak symmetry
$SU_L(2) \times U_Y(1)$ is broken down to the electromagnetic symmetry
$U_{em}(1)$.
As a result, fermions get masses and three Goldstone bosons appear.
One can see that there are Goldstone-Fermion vertices from the Yukawa
sector. To see it clearly, let us parametrize $\Phi$ as
$${
\Phi =  \Sigma ~ \pmatrix{ 0
\cr {v + H} \over {\sqrt 2} \cr } \simeq {1 \over {\sqrt 2}}
         \pmatrix{  \xi^2 + ~ i \xi^1 \cr v + H - i \xi^0 \cr }
      ~~~~, }\eqno(2)$$
 \no where $\Sigma = e^{i {{\vec \xi} \cdot {\vec \tau} \over v} }$,
$H$ is the physical Higgs boson and
$\xi^i$ the Goldstone particles. From
${\cal L}_3$, one obtains the ``linear" lagrangian,
which describes the non-derivative
interactions of the fermions to the Goldstone bosons. To get the
 ``non-linear" lagrangian, one usually redefines the fermion fields
as follows,
$${
  {\pmatrix{t^\prime \cr b^\prime \cr }}_L
= ~ \Sigma ~ {\pmatrix{t \cr b \cr }}_L
{}~,  ~t^\prime_R = t_R ~, ~{\rm and } ~~ b_R^\prime = b_R ~ ~. }\eqno(3)$$
\no In the ``nonlinear" lagrangian
 the Goldstone bosons disappear from the Yukawa
 sector in ${\cal L}_3$, but come out from the kinetic energy term in
${\cal L}_1$. As a result, the Goldstone bosons interact with the
fermions with derivatives, which is given explicitly by
$${ \eqalign{
{\cal L}^{S.M.} = &  {\overline \pmatrix{t \cr b \cr } }_L \Sigma^\dagger ~
i \gamma^\mu \partial_\mu ~ \Sigma { \pmatrix{ t \cr b \cr } }_L  \cr
    = &  -{1 \over v}~ ( {\overline t_L} \gamma^\mu t_L - {\overline b}_L
  \gamma^\mu b_L )~ \partial_\mu \pi^0   \cr
     & - {\overline t_L} \gamma^\mu b_L ~ \{ ~
{{\sqrt 2}\over v} \partial_\mu \pi^+
    ~ + ~i~ ( {{\sqrt 2} \over v^2} ) ~[ ~ \pi^+ \partial_\mu \pi^0
             - \pi^0 \partial_\mu \pi^+ ~]~  \} \cr
     & + h.c. ~ \cr
     & + ...... \cr}
} \eqno(4)$$

\no where $\pi^0 \equiv \xi^3 ~, ~ \pi^+ \equiv {1 \over {\sqrt 2} }~ ( \xi^1
              - i \xi^2) $.
The ``linear" and ``nonlinear" lagrangians
generate the same S-matrix elements[4]. However, it proves more convenient
to use the fermion basis in (3) for the ``nonlinear"
lagrangian in
studing the low energy effects of the heavy Higgs boson, chiral
fermions and definitly in the construction of the
effective lagrangian of the dynamical symmetry breaking theory.

 These Goldstone fields,
after gauging the electroweak symmetry,
 become the longitudinal gauge bosons.
The point we want to stress here
is that the universal coupling of the
matter fields to the Goldstone bosons
( for example, $\pi^0$ couples to the third component of the $SU(2)$
current with strength equally for top and bottom quarks, as well as
other fermions, ) results in
 the universality of the
weak interaction.
  In the standard model, however,
the Yukawa interaction is not universal,
so one would expect to have non-universal corrections at
quantum level.
This kind of loop effect is important for
 the ${\pi^0} - {\overline b}- b$ vertex because of the heavy top quark.

Generally the effective vertex $\pi^0 -{\overline b} -b$ can be expressed
\footnote{[F.1]}{
After gauging $SU_L(2) \times U_Y(1)$, and going into
unitary gauge, ${\cal L}_b \rightarrow {g \over {2~ \cos\theta_W} }
 \hfil\break
( 1 + \tau_b ) ~
{\overline b}_L \gamma^\mu b_L ~Z_\mu$ . } as
$${
{\cal L}_b = ( 1 + \tau_b ) ~( {1\over v} )~
           {\overline b}_L \gamma^\mu b_L
                            \partial_\mu \pi^0
 ~+ ~ O( {1\over v^2} )~~~~, }\eqno(5)$$
\no where $\tau_b$ is a parameter to characterize the non-universal effect.
We should exphasize that the effective operator in ${\cal L}_b$ is
$SU_L(2) \times U_Y(1)$ invariant. It can be clearly seen in the ``nonlinear"
lagrangian, where the matter fields only directly feel the electromagnetic
charge and interact with Goldstone bosons with derivative.
Let us now calculate the
 $\tau_b$ in the lagrangian (4).
The corresponding
 Feynman diagrams are shown in Fig.1.
 The integration given by each of these
diagrams is divergent, but their sum is finite.
 This is expected since the lagrangian (4) is derived from
renormalizable lagrangian (1).
Taking the external bottom
quark massless, our one-loop result is
$${
\tau_b^{S.M.} = - {2 \over { 16 \pi^2 } } {m_t^2 \over v^2 }
 ~~~~. } \eqno(6)$$

\no This is a well-known result[5]. However, the calculation here is
simpler than that generally used with gauge fields included. In
fact, it will become much more simpler in the ``linear" lagrangian,
where there is only one diagram instead of three in the ``nonlinear"
 lagrangian\footnote{[F.2]}{
Recently R. Barbieri et al[6] have used the ``gauge-less"
lagrangian, the lagrangian of
the standard model in the limit of vanishing gauge coupling
constants,
 to calculate the $\rho$ and
$\tau_b$ up to two loop.
 They used ``linear" lagrangian. Our one-loop result
using the ``non-linear" lagrangian agrees with their result.}
. But, as we pointed out above, the ``nonlinear"
lagrangian is suitable for the dynamical symmetry breaking theory.
In this paper we will focus
 on the calculation of $\tau_b$ in the chiral lagrangian of the
dynamical
symmetry breaking theory.

In the dynamical symmetry breaking theory, the vertices of the
top quark with the Goldstone bosons are expected to be
different from
that of the standard model.
In addition to the terms in (4), there are some
new interactions. In the notation of
Ref.[1], these new interactions can be parametrized as follows,
$${ \eqalign{
{\cal L}_{new} \simeq &  -  ~ (\kappa_L^{NC})~
         {1 \over v}~ {\overline t}_L \gamma^\mu t_L ~
 \partial_\mu \pi^0~  - ~ ( \kappa_R^{NC}) ~{1\over v} ~
            {\overline t}_R \gamma^\mu t_R   \partial_\mu
             \pi^0   \cr
           & - (\kappa_L^{CC}) ~ {\overline t}_L \gamma^\mu b_L~
              \{ ~
{{\sqrt 2}\over v}~ \partial_\mu \pi^+
                ~  +~ i ~{ {\sqrt 2} \over v^2 }~ [ \pi^+ \partial_\mu \pi^0
                      - \pi^0 \partial_\mu \pi^+ ]~ \} \cr
           & - (\kappa_R^{CC}) {\overline t}_R \gamma^\mu b_R~~
               \{ ~ {{\sqrt 2}\over v}
          \partial_\mu  \pi^+
            ~  +~  i ~ { {\sqrt 2} \over v^2 } ~ [ \pi^+ \partial_\mu \pi^0
                        - \pi^0 \partial_\mu \pi^+ ]~  \} ~~ \cr
           & + ~ h.c.     \cr }
   } \eqno(7)$$

\no After gauging $SU_L(2) \times U_Y(1)$, the
 $\kappa_{L(R)}^{NC}$ term gives the new neutral current interaction for
 left-handed (right-handed) top quarks,
 and $\kappa_{L(R)}^{CC}$ term
is the new
 left-handed (right-handed) charged current interaction[1].

 Let us calculate again
 the top loop contributions to $\tau_b$ in the presence of ${\cal L}_{new}$.
Since the new interactions can generally be treated as
 ``small perturbative" corrections to the
standard model,
 we need keep only the linear terms in
 $\kappa$ in the calculation.
 This leads us to reexamine the diagrams in Fig.1
 with replacing only one
of the vertices by
the corresponding $\kappa$ term in ${\cal L}_{new}$. The chiral
 lagrangian in Ref.[1] is not renormalizable in the commen sense. It needs
a term, which is similar to the $\kappa_L^{NC}$
in eq.(7), for the bottom quark as a
counterterm. Clearly this counterterm
 will give rise to a direct contribution to the
$\tau_b$. In our calculation we use the
dimensional regularization
to regularize the
integral in order to preserve the chiral invariance of
 the lagrangian, and take the renormalization scale, $\mu$, to be
$\Lambda$, the cut-off of the effective lagrangian.
 Following these rules, we obtain
 that\footnote{[F.3]}{The term proportional to
$\kappa_L^{NC}$ and
$\kappa_R^{NC}$ was obtained in Ref.[7] by doing the calculation in
the unitary gauge.
Here we recover this result in an easy way.
}
$${
\eqalign{
{\tau_b^{new}} = & \tau_b^0(  \Lambda  ) \cr
                     & + { {4 ~ \kappa_L^{NC} - \kappa_R^{NC} }
 \over {16 ~ \pi^2 } } ~ {m_t^2 \over v^2} ~ \ln { \Lambda^2 \over m_t^2 } \cr
                    & + { 4 \kappa_L^{CC} \over {16~ \pi^2 }}
                       {m_t^2 \over v^2 } \ln { \Lambda^2 \over m_t^2 }
                         ~~~~~. \cr }
}\eqno(8)$$

\no A few remarks are now in order:

\item{i)} $\tau_b^0 (  \Lambda  )$ is the counterterm contribution.
In the chiral lagrangian,
it is totally a free parameter.
However once the
underlying theory for the symmetry breaking is known,
 $\tau^0_b( \Lambda )$ can be calculated principally.
 For instance, in pure technicolor theories it vanishes
 since the quarks and leptons are totally decoupled from
the symmetry breaking sector. In extended technicolor (ETC) theories, the
$\kappa$ terms are induced {\it via} the ETC interactions[1]
\footnote{[F.4]}{ The non-universal effects in the $Z ~b ~ {\overline b}$
vertex
from ETC dynamics
has been recently studied in detail in Refs.[8]. }.

\item{ii)}There is no contribution from $\kappa_R^{CC}$.
This result can also be understood qualitatively as followes.
Since $\kappa_R^{CC}$ term is a right-handed, and others are
the left-handed interaction,
the operator, which may be generated at one-loop,
should be proportional to
$i ~ {\overline b}_L~ b_R ~ \pi^0$
in the limit $m_b = 0$ . This operator, however, breaks the
chiral invariance, then is forbidden.

\no Putting the results in eqs.(6)
and (8) together, we have

$${ \eqalign{
\tau_b =&  ~ \tau_b^{S.M.} ~+ ~ \tau_b^{new} \cr
       = & - {2 \over { 16 \pi^2 } }~ { m_t^2 \over v^2 } \cr
        & + \tau^0_b( \Lambda ) \cr
        & + { 4 \kappa_L^{NC} - \kappa_R^{NC} \over { 16~ \pi^2 }}
           ~ { m_t^2 \over v^2 } ~ \ln {\Lambda^2 \over m_t^2 } \cr
        & +{ 4~ \kappa_L^{CC} \over { 16 ~ \pi^2 }} ~ {m_t^2 \over v^2 }
           ~ \ln { \Lambda^2 \over m_t^2 } ~~~~. \cr }
  } \eqno(9)$$

\no We should exphasize that what we have calculated here are
 the corrections to $Z b {\overline b}$ vertex which are not
suppressed by the gauge coupling $\alpha_W$. If
including weak gauge bosons in the chiral lagrangian,
 there are other contributions to $\tau_b$
indicated, for example, by the Feynman
diagram in Fig.2\footnote{[F.5]}{
This loop effect on the $Z b {\overline b}$
partial width, even though
suppressed by $\alpha_W$,
can be used to put constraints on the anomalous
triplet gauge boson couplings from
the LEP data[9].}.
 The triplet Goldstone boson vertices, however,
in the chiral lagrangian are forbbiden
if the symmetry breaking sector preserves an $SU(2)_{L+R}$ symmetry
 ( $\rho = 1$ ) .
If $\delta \rho \not= 0$, there may be contributions to $\tau_b$ from
the triplet Goldstone bosons vertices. But
 it is at most
$O( ~\alpha_W ~ )$ since $\delta \rho$ is less than
$\alpha_W$.

To estimate the value of $\tau_b^{new}$, we need to know the size
of $\tau^0_b (\Lambda )$
and all of the $\kappa$ in eq.(7). They are free parameters in the
chiral lagrangian. However, it is quite natural[1]
to assume that a certain correlation with the fermion mass may exist, with
the large effects occuring only for heavier particles. Thus, logically,
 the neutral
current interactions
 of the top quark is one of the most natural place
to search for these new physics effects. In the following numerical evaluation
of the
$\tau_b^{new}$, we shall keep only the $\Delta g_A = g_A -1$ term,
where $g_A$ is the axial vector coupling constant, for the
neutral current interaction of the top quark
with the neutral gauge boson Z. The effective operator
responsible for $\Delta g_A$ in the chiral lagrangian is given by
$${
{\cal L}_{\Delta g_A} = { (\Delta g_A) \over v} ~{1\over 2} ~{\overline t}
       \gamma^\mu \gamma_5 t ~ \partial_\mu \pi^0 ~+ ~ O( {1 \over v^2} )
{}~~~. } \eqno(10)$$
\no This
corresponds to that
$\Delta g_A =  2~ \kappa_L^{NC}$ in eq.(7) by
 setting $\kappa_L^{NC} = - \kappa_R^{NC}$.
 From eq.(8), one obtains the
contribution of $\Delta g_A$ of the top quark to $\tau_b$,
$${
\tau_b^{\Delta g_A} =  {5 ~ (\Delta g_A) \over {32~ \pi^2 }} {m_t^2 \over v^2}
\ln {\Lambda^2 \over m_t^2}   ~~~~~. }  \eqno(11)$$
\no Now $\tau_b$ in (9) is reduced to
$${
\tau_b = - {2 \over {16 ~ \pi^2 }} {m_t^2 \over v^2 }
        + ~({\Delta g_A})~ { 5 \over {32 ~\pi^2 } }{m_t^2 \over v^2}
          \ln { \Lambda^2 \over m_t^2 } ~~~~. } \eqno(12)$$

\no Clearly a positive $\Delta g_A$ can make
the theoretical prediction more comparable to the LEP data
than a negative
$\Delta g_A$ does[10]. To illustrate it we plot the $\tau_b$ as function
 of the top quark masses in Fig.3, for
$\Delta g_A = 0.25$ and
$\Lambda = 1$ TeV. The arrows in the Fig.3 indicate the 1 $\sigma$
data on $\tau_b$, which is taken from the first paper of Ref.[3]~
[ $\epsilon_b = ( 3.0 \pm 5.8 ) ~ 10^{-3}$ ], obtained by a four-parameter
fit, in terms of $\epsilon_1, \epsilon_2, \epsilon_3 ~{\rm and}~ {\epsilon_b}$
using the data on $\Gamma_l, A_{FB}^l, m_W/m_Z ~{\rm and}~ \Gamma_b$ for
$\alpha_s( m_Z ) = 0.118$.

 In conclusion, we have studied systematically
the non-universal effects on the $Z b {\overline b}$ vertex in the dynamical
electroweak symmetry breaking theory.
 We have calculated $\tau_b$ at one loop in the chiral
lagrangian which includes only the
fermions and the Goldstone bosons. Since
$\tau_b$ in (9) includes
  the corrections which are not
suppressed by the gauge coupling constants,
 it reflectes directly the
physics associated with the heavy top quark
and the mass generation in the dynamical symmetry breaking
theory.
\bb

\vfill\eject
\b
I would like to thank G. Valencia and B.-L. Young for discussions.
 This work is supported in part by the
Office of High Energy and Nuclear Physics of the U.S. Department of
Energy (Grant No. DE-FG02-94ER40817).
\b

\ce {\bf REFERENCES}

\item{[1]}R.D. Peccei and X. Zhang, Nucl. Phys. B337, 269 (1990).

\item{[2]}M. Goldberger and S. Treiman, Phys. Rev. V110, 1178 (1958).

\item{[3]}see, for examples,
G. Altarelli, CERN-TH 7072/93;
D. Comelli, C. Verzegnassi and F. Renard,
             PM/94-04, UTS-DFT-94-04; and references therein.

\item{[4]}S. Coleman, J. Wess and B. Zumino, Phys. Rev. 177, 2239 (1969);
         C.G. Callan, S. Coleman, J. Wess and B. Zumino, Phys. Rev. 177
        2247 (1969).

\item{[5]}A. Akhundov, D. Barlin and T. Riemann, Nucl. Phys. B276, 1 (1986);
               J. Bernabeu, A. Pich and A. Santamaria, Phys. Lett. B200,
      569 (1988); W. Beenakker and W. Hollik, Z. Phys. C40, 141 (1988);
              F. Boudjema, A. Djuoadi and C. Verzegnassi, Phys. Lett.
        B238, 423 (1990).

\item{[6]}R. Barbieri, M. Beccaria, P. Ciafaloni, G. Curci and A. Vicer\'e ,
          Phys. Lett. B288, 95 (1992); Nucl. Phys. B409, 105 (1993).

\item{[7]}R.D. Peccei, S. Peris and X. Zhang, Nucl. Phys. B349, 305 (1991).

\item{[8]}R.S. Chivukula, S.B. Selipsky and E.H. Simmons, Phys. Rev. Lett.
        69, 575 (1992); R.S. Chivukula, E.H. Simmons and J. Terning,
BUHEP-94-8, hep-ph/9404209; N. Kitazawa, Phys. Lett. B313, 395 (1993);
N. Evans, University of Wales, Swansea Preprint SWAT/27, hep-ph/9403318.

\item{[9]}S. Dawson and G. Valencia, (unpublished).

\item{[10]}Using LEP data
 to constrain the $\kappa$ in (7) can
be found in Ref.[7]; and, M. Frigeni and R. Rattazzi, Phys. Lett.
B269, 412 (1991); and, D.O. Carlson, E. Malkawi and C.-P. Yuan,
Michigan State University, MSUHEP-94/05, May (1994).

\bye
\bye